\documentclass[a4paper,12pt]{article}
\usepackage{amsmath,amsthm,amssymb,bbm}

\usepackage[english]{babel}
\usepackage{graphicx}

\title{\bf Environment induced entanglement in many-body mesoscopic systems}

\author{F. Benatti$^{a,b}$, F. Carollo$^{a,b}$
R. Floreanini$^{b}$\\
\\
\small ${}^a$Dipartimento di Fisica, Universit\`a di Trieste, 
34151 Trieste, Italy\\
\small ${}^b$Istituto Nazionale di Fisica Nucleare, Sezione di Trieste,
34151 Trieste, Italy}
\date{\null}

\begin{document}

\maketitle

\begin{abstract}
\noindent
We show that two, non interacting, infinitely long spin chains
can become globally entangled at the mesoscopic level
of their fluctuation operators through a purely noisy microscopic 
mechanism induced by the presence of a common heat bath. By
focusing on a suitable class of mesoscopic observables,
the behaviour of the dissipatively generated quantum correlations
between the two chains is studied as a function of the
dissipation strength and bath temperature.
\end{abstract}

\vskip 1cm

The presence of an external environment, typically a heat bath, 
modifies the dynamics of quantum systems in interaction with it,
leading in general to loss of quantum correlations
due to decohering and mixing-enhancing effects \cite{Alicki}.
Nevertheless, it has also been established that suitable
environments can enhance quantum entanglement 
instead of destroying it \cite{Benatti1}.

This mechanism of environment induced entanglement generation
has been extensively studied for systems made of few qubits 
or oscillator modes \cite{Benatti2}, and specific protocols have been proposed
able to prepare predefined entangled states via the action of suitably
engineered environments \cite{Kraus}.

Instead, in this paper, we study the possibility that entanglement be created 
through a purely noisy mechanism in many-body systems.
In a quantum system made of a large number $N$ of constituents, 
accessible observables are collective ones, {\it i.e.} those
involving the degrees of freedom of all its elementary parts.
For these ``macroscopic'' observables, one usually expects that quantum effects 
fade away as $N$ becomes large, even more so when
the many-body system is in contact with an external environment.
This is surely the case for the so-called ``mean field'' observables,
{\it i.e} averages over the whole system of microscopic
operators; these quantities scale as $1/N$
and as such behave as classical observables when the number
of system constituents becomes large.

Nevertheless, other collective observables exist that scale as
$1/\sqrt{N}$ and that might retain some quantum properties
as $N$ increases \cite{Goderis,Verbeure,Matsui}. These observables have been called
``fluctuation operators'', since, as we shall see, they physically represent
some sort of deviations from mean values.
The set of all these fluctuation operators form an algebra that, irrespective
of the nature of the microscopic many-body system, turns out to be always
non commutative and of bosonic type, thus showing
a quantum behaviour. Being half-way between microscopic
observables (as for instance the individual spin operators 
in a generic spin systems) and truly macroscopic ones 
({\it e.g.} the corresponding mean magnetization),
the fluctuation operators have been named ``mesoscopic'':
they are the place where to look for truly quantum signals
in the dynamics of ``large'' systems, {\it i.e.} in systems 
in which the number of microscopic constituents is let arbitrarily growing
at fixed density (thermodynamical limit).

Although the characteristics and time evolution of the fluctuation algebra have 
been extensively studied in many systems \cite{Verbeure}, very little is known
of its behaviour in open many-body systems, {\it i.e.} in systems 
immersed in an external bath.
This is the most common situation encountered in actual experiments, typically
involving cold atoms, optomechanical or spin-like systems \cite{Aspelmeyer,Rogers}, that can
never be thought of as completely isolated from their thermal surroundings.
Actually, the repeated claim of having detected ``macroscopic'' entanglement
in those experiments \cite{Jost,Krauter} poses a serious challenge in trying to 
interpret theoretically those results \cite{Narnhofer1,Narnhofer2}.

Motivated by these experimental findings, in the following we shall show
that quantum behaviour can indeed be present at the mesoscopic level in open many-body
systems provided suitable fluctuation operators are considered and,
even more strikingly, that entanglement can be induced in mesoscopic observables
by the presence of the external bath.

We shall consider a many-body system composed by two spin-1/2 chains, one next to the other,
immersed in a heat bath at a given inverse temperature $\beta=1/T$. Each site of this double chain,
actually composed by the corresponding couple of sites in the two chains,
will be labelled by an integer $k=1,2,\ldots, N$. In this situation, the thermodynamical limit corresponds
to letting the total number of sites $N$ going to infinity.

A spin algebra $\cal M$, corresponding
to the tensor product of two spin-1/2 algebras, is attached to each site; its elements at site $k$,
$x^{(k)}\in {\cal M}^{(k)}$, are then
of the form $x^{(k)}=x_1^{(k)}\otimes x_2^{(k)}$, where $x_1^{(k)}$, $x_2^{(k)}$ are
spin algebra elements pertaining to the first, second chain, respectively.
The algebra $\cal M$ clearly coincides with the algebra of $4\times 4$ complex matrices
and a convenient basis in it is given by $\sigma_\mu\otimes \sigma_\nu$, $\mu,\nu=0,1,2,3$,
where $\sigma_i$, $i=1,2,3$ are the usual Pauli matrices, while $\sigma_0$ is the unit matrix.
For any finite set $I$ of contiguous sites, one defines the finite-size tensor algebra 
${\cal A}_I=\otimes_{k\in I} {\cal M}^{(k)}$; the union $\cal A$ of all these algebras,
${\cal A}=\cup_I  {\cal A}_I$,
is called the {\sl quasi local algebra} and the observables of the system clearly belong to it.

A state $\omega$ for the system is a linear, positive, normalized functional 
on the algebra $\cal A$, $\omega: {\cal A} \to \mathbb{C}$, assigning the
expectation value $\omega(X)$ to each elements $X$ of $\cal A$.
For finite $N$, it can be represented by a density matrix $\rho$ through
the identification $\omega(X)={\rm Tr}[\rho\, X]$; however, since we are interested in the
thermodynamical limit, it is more convenient to work in the abstract
algebraic formulation \cite{Bratteli}. 

Since the two chains can be thought to be initially at equilibrium with the bath,
as reference state for our system we take a product state
\begin{equation}
\omega= \omega^{(1)}\otimes\omega^{(2)}\otimes\omega^{(3)}\otimes\ldots\ ,
\label{1}
\end{equation}
where $\omega^{(k)}$, $k=1,2,3,\ldots$, are single site states, that for simplicity 
can be assumed to be all equal to a reference thermal state, at the bath temperature.
As a consequence, $\omega$ has the property that given two observables $x^{(k)}$, $y^{(l)}$
at different sites, $k\neq l$, then: $\omega(x^{(k)}\, y^{(l)})=\omega(x^{(k)})\, \omega(y^{(l)})$;
this means that in practice $\omega$ is uniquely defined by the expectation values 
on all observables \hbox{$x\in{\cal M}$} at one site of the double chain, 
that in the following will be simply called $\omega(x)$.

Most of the physical properties of many-body systems can be obtained by focusing 
on collective observables, {\it i.e.} on operators involving all system degrees of freedom, which,
in the present situation, means combinations of spin variables at all $N$ sites.
In the thermodynamical limit, {\it i.e.} when $N$ becomes infinitely large, a suitable
scaling with $N$ needs to be included in the definition of these observables
in order to obtain meaningful limiting operators. 

A well-known example of such observables is given by the averages over all sites of
a given spin operator $x\in {\cal M}$:
\begin{equation}
\overline{X}_N=\frac{1}{N} \sum_{k=1}^N x^{(k)}\ .
\label{2}
\end{equation}
As $N$ grows, the sequence of operators $\{\overline{X}_N\}$ converges to the ``macroscopic'' observable
$\overline{X}=\lim_{N\to\infty} \overline{X}_N$. This convergence should be intended in the
weak sense, {\it i.e.} under state average.%
\footnote{More precisely, weak convergence means that, given any couple of local
operators $Y$ and $Z$ having support only on a finite number of sites, the sequence 
$\omega(Y\, \overline{X}_N\, Z)$ converges in the limit of large $N$;
because of the assumed form of the state $\omega$, one further has:
$\lim_{N\to\infty}\omega(Y\, \overline{X}_N\, Z)=\omega(YZ)\, \omega(x)$,
and thus $\lim_{N\to\infty} \overline{X}_N=\omega(x)\, {\bf 1}$.}
In the case of the product state (\ref{1}), this limit is easily computed:
$$
\lim_{N\to\infty} \omega(\overline{X}_N)=\lim_{N\to\infty}\frac{1}{N}\sum_{k=1}^N \omega(x^{(k)})=\omega(x)\ ,
$$
since the expectations $\omega(x^{(k)})$ are all equal and independent of $k$. 
In practice, one obtains \cite{Goderis}
\begin{equation}
\overline{X}=\lim_{N\to\infty} \overline{X}_N=\omega(x)\, {\bf 1}\ ,
\label{3}
\end{equation}
with ${\bf 1}$ the identity operator. As a result,
the set of all these limiting operators form an abelian algebra, since all operators commute among themselves;
it is called the {\sl mean field} algebra and it is known to represent the classical behaviour of the system.

Nevertheless, some of the system quantum properties can survive even in the large $N$ limit: 
they are encoded in the so-called {\sl fluctuation operators}. These are collective
observables that scale as the square root of $N$,
\begin{equation}
\widetilde{X}=\lim_{N\to\infty} \widetilde{X}_N\equiv 
\lim_{N\to\infty}\frac{1}{\sqrt N} \sum_{k=1}^N \Big[x^{(k)}-\omega(x^{(k)})\Big]\ ,
\label{4}
\end{equation}
and represent a sort of deviation from (or fluctuation about) the average.
One easily sees that the commutator of two such fluctuation observables is in general nonvanishing,
since it is equal to a mean field operator
\begin{equation}
[\widetilde{X},\, \widetilde{Y}]=\lim_{N\to\infty} [\widetilde{X}_N, \, \widetilde{Y}_N]
=\lim_{N\to\infty}\frac{1}{N}\sum_{k=1}^N\, [x^{(k)}, \, y^{(k)}]\ ,
\label{5}
\end{equation}
being $[x^{(k)}, \, y^{(l)}]=\,0$ for $k\neq l$. Recalling (\ref{3}), this implies that
$[\widetilde{X},\, \widetilde{Y}]$ is proportional to the identity operator,
and therefore that the algebra formed by all fluctuation operators possesses a quantum character,
being non-abelian.

The fluctuation algebra is clearly bosonic and look very similar to the Heisenberg 
algebra of position and momentum operators; as in that case, the algebra elements $\widetilde{X}$
in (\ref{4}) turn out to be unbounded operators,
their norm diverging as $\sqrt N$ in the thermodynamical limit. To avoid convergence problems,
it is then convenient to work with the corresponding Weyl operators, 
$\lim_{N\to\infty} e^{i \widetilde{X}_N}$,
whose existence in the weak sense is guaranteed 
by the so-called {\sl quantum central limit} \cite{Goderis,Verbeure,Matsui}.
Indeed, defining the following sesquilinear form on the algebra of fluctuations:
\begin{equation}
\langle\widetilde{X},\, \widetilde{Y}\rangle_\omega=
\lim_{N\to\infty}\omega\big( \widetilde{X}_N^\dagger \, \widetilde{Y}_N\big)\ ,
\label{6}
\end{equation}
one shows that, for any hermitian spin operator $x$, the following result holds:
\begin{equation}
\lim_{N\to\infty}\omega\big( e^{i \widetilde{X}_N} \big)=
e^{-\frac{1}{2} \langle\widetilde{X},\, \widetilde{X}\rangle_\omega}\ .
\label{7}
\end{equation}
Similarly, products of any number of Weyl operators can be analogously computed;
in particular, one has
\begin{equation}
\lim_{N\to\infty}\omega\big( e^{i \widetilde{X}_N}\, e^{i \widetilde{Y}_N} \big)=
e^{-\frac{1}{2}\big( \langle\widetilde{X}+\widetilde{Y},\, \widetilde{X}+\widetilde{Y}\rangle_\omega
+ [\widetilde{X},\, \widetilde{Y}]\big)}\ ,
\label{8}
\end{equation}
with
\begin{equation}
[\widetilde{X},\, \widetilde{Y}]=
2i\, {\cal I}m\big( \langle\widetilde{X},\, \widetilde{Y}\rangle_\omega \big)\, {\bf 1}\ .
\label{9}
\end{equation}
In other terms, in the large $N$ limit, the set of hermitian fluctuation operators 
$\{\widetilde{X}\}$ form a well defined bosonic algebra, characterized by the
commutation relations (\ref{9}). This algebra can be appropriately described in terms of
the Weyl operators $W(x)=e^{i \widetilde{X}}$ and 
a suitable Gaussian state $\widetilde{\omega}$ reproducing all higher order correlations:%
\footnote{Given a state $\omega$ over an algebra of operators $\cal A$, a standard procedure,
the so-called {\sl GNS construction} \cite{Bratteli}, allows to build an Hilbert space ${\cal H}_\omega$,
generated by a cyclic ``vacuum'' vector $|\Omega_\omega\rangle$,
and a representation $\pi_\omega$ of $\cal A$ into the bounded operators on ${\cal H}_\omega$;
further, the expectation of any element $X\in{\cal A}$ 
is given by the corresponding vacuum mean value, {\it i.e.}
$\omega(X)=\langle\Omega_\omega| \pi_\omega(X) |\Omega_\omega\rangle$. Therefore, the Weyl correlations
in (\ref{10}) reproduce the mean value of any fluctuation observable in any state of the 
corresponding Hilbert space.}
\begin{equation}
\widetilde{\omega}\big(W(x)W(y)\ldots\big)=
\lim_{N\to\infty}\omega\big( e^{i \widetilde{X}_N}\, e^{i \widetilde{Y}_N}\ldots \big)\ .
\label{10}
\end{equation}
We want now to study the dynamics and its possible entanglement generating properties on
the states of this algebra.

As mentioned before, the two chains are assumed to be non interacting, 
so that the free system Hamiltonian at finite $N$ will be taken to be of the simple form
\begin{equation}
H_N=\sum_{k=1}^N H^{(k)}=
\frac{\epsilon}{2}\,\sum_{k=1}^N \big(\sigma_3^{(k)}\otimes\sigma_0 
+ \sigma_0\otimes\sigma_3^{(k)}\big)\ ,
\label{11}
\end{equation}
with $\epsilon$ a constant energy parameter. Being initially at equilibrium with the bath, 
the state $\omega$ for the system can be written 
as a product of single site thermal states, see (\ref{1}), with 
$\omega^{(k)}(\cdot)={\rm Tr}[e^{-\beta H^{(k)}} \cdot\,]/{\rm Tr}[e^{-\beta H^{(k)}}]$.

Because of the presence of the bath, the dynamics of any observable $X$ of the system is not
generated by $H_N$ alone; additional pieces accounting for dissipative and noisy effects will be present.
For a weakly coupled bath, using standard techniques \cite{Alicki,Benatti2}, 
a Markovian equation of Kossakowski-Lindblad form can be derived:
\begin{equation}
\frac{d X_t}{d t}= i [H_N,\, X_t] + \mathbb{L}_N[X_t]\equiv {\cal L}_N[X]\ .
\label{12}
\end{equation}
Assuming the bath to be coupled in the same way to all sites,
the dissipative part of the generator can be written as \cite{Alicki,Benatti2}:
\begin{equation}
\mathbb{L}_N[X]=\sum_{k,l=1}^N J_{kl}\sum_{\mu,\nu=1}^4 
D_{\mu\nu}\big[[V_\mu^{(k)},X],V_{\nu}^{(l)\dagger}\big]\ ,
\label{13}
\end{equation}
where $\{V_\mu^{(k)}\}_{\mu=1}^{4}$ are the following two-chain operators
$\{\sigma_+\otimes \sigma_-,\, \sigma_-\otimes \sigma_+,
\, \sigma_3\otimes \sigma_0/2,\,$\break $\sigma_0\otimes \sigma_3/2\}$
at site $k$, with $\sigma_{\pm}=(\sigma_1\pm i\sigma_2)/2$. The matrix $D$
encodes the bath noisy properties; it will be taken of the following non-diagonal form,
depending for simplicity on a single dissipative parameter $\gamma$:
\begin{equation}
D=
\begin{pmatrix}
1&0&\gamma&\gamma\\
0&1&\gamma&\gamma\\
\gamma&\gamma&1&0\\
\gamma&\gamma&0&1
\end{pmatrix}\ .
\label{14}
\end{equation}
On the other hand, the coefficients $J_{kl}$ take care of possible couplings between the sites.%
\footnote{It is reasonable to take $J_{kl}$ to be translationally invariant and such that
$\sum_{k,l=1}^N |J_{kl}|\simeq N$.}
The condition of complete positivity, essential for assuring the physical consistency of the dynamics
generated by (\ref{12}) \cite{Benatti2}, requires $[J_{kl}]\geq0$ and $\gamma\leq1/2$. 
Furthermore, notice that the four Lindblad
operators $V_\mu^{(k)}$ commute with $H^{(k)}$; therefore, the chosen thermal state $\omega$ is left invariant
by the finite time dynamics generated by the equation (\ref{12}): $\omega \circ e^{t {\cal L}_N}=\omega$.

Although the spin algebra $\cal M$ attached to each site of the double-chain system is 16-dimensional,
we find convenient to focus on the following subset of eight observables that we orderly label
$x_\alpha$, $\alpha=1,2,\ldots,8$,
\begin{equation}
\begin{matrix}
\sigma_1\otimes\sigma_0\ , & \sigma_2\otimes\sigma_0\ , & 
\sigma_1\otimes\sigma_3\ , & \sigma_2\otimes\sigma_3\ ,\\
\sigma_0\otimes\sigma_1\ , & \sigma_0\otimes\sigma_2\ , & 
\sigma_3\otimes\sigma_1\ , & \sigma_3\otimes\sigma_2\ ,
\end{matrix}
\label{15}
\end{equation}
and the corresponding fluctuation operators $\widetilde{X}_\alpha$ defined as in (\ref{4}).%
\footnote{One can show that the remaining eight independent observables in $\cal M$, complementary to those listed
in (\ref{15}), give rise to fluctuation operators that in the thermodynamical limit commute with the set
$\{\widetilde{X}_\alpha\}$.}
By defining the combinations (for $\beta$ finite):
\begin{eqnarray}
&& a_1=\frac{1}{2\, \sqrt\eta}\big[ \widetilde{X}_1 -i \widetilde{X}_2 \big]\ ,\\
&& a_2=\frac{\sqrt\eta}{2\,\sqrt{1-\eta^2}}\Big[\widetilde{X}_1 -i \widetilde{X}_2
+\frac{1}{\eta} \big( \widetilde{X}_3 -i \widetilde{X}_4\big) \Big]\ ,
\end{eqnarray}
with $\eta=\tanh(\epsilon\beta/2)$, 
and the two additional ones $b_1$ and $b_2$ obtained from the previous ones with the substitution
$\widetilde{X}_\alpha \to \widetilde{X}_{\alpha+4}$, one easily shows that the fluctuation operators
obey standard canonical commutation relations:
\begin{equation}
[a_i,\, a_j^\dagger]=\delta_{ij}\ ,\quad [b_i,\, b_j^\dagger]=\delta_{ij}\ ,\quad i,j=1,2\ ,
\label{18}
\end{equation}
while all other commutators vanish.%
\footnote{In the limit of zero temperature, the algebra of fluctuation contracts, and only
the combinations $a_1$ and $b_1$ survive.}
Furthermore, the Gaussian state that reproduce all
the large $N$ Weyl correlations as in (\ref{10}) is again a thermal state, explicitly given by 
$\widetilde{\omega}(\cdot)={\rm Tr}[e^{-\beta\widetilde{H}}\, \cdot\,]/{\rm Tr}[e^{-\beta\widetilde{H}}]$,
where $\widetilde{H}=\epsilon\sum_{i=1}^2 \big(a_i^\dagger a_i + b_i^\dagger b_i \big)$ is the
free Hamiltonian for the fluctuation modes.

The time evolution of the fluctuation operators $\widetilde{X}_\alpha$ in the large $N$ limit
is induced by the microscopic one at finite $N$ generated 
by the Master Equation (\ref{12}):  $\lim_{N\to \infty} e^{t {\cal L}_N}[ e^{i\widetilde{X}_N}]
=e^{t \widetilde{\cal L}}[e^{i\widetilde{X}}]$. It can be conveniently expressed
as acting on the Weyl operators of the four modes $a_i$, $b_i$, 
defined in the standard way as ($z_\mu\in\mathbb{C}$, $\mu=1,2,3,4$)
\begin{equation}
W(z)=e^{A(z)^\dagger-A(z)}\ ,\quad A(z)=\sum_{i=1}^2 \big(z_i a_i +z_{2+i} b_i\big)\ .
\label{19}
\end{equation}
One can then show that the induced time evolution is of {\sl quasi-free} type \cite{Verbeure}, 
{\it i.e.} mapping Gaussian states into Gaussian states; it is explicitly given by \cite{Benatti3}:
\begin{equation}
e^{t \widetilde{\cal L}}\big[W(z)]=e^{\varphi(t)}\, W(z(t))\ ,
\label{20}
\end{equation}
where the evolution of the four-vector $|z(t)\rangle$ with components $z_\mu(t)$, $\mu=1,2,3,4$,
is linear, $\frac{d}{dt}|z(t)\rangle=M|z(t)\rangle$, with initial condition $z_\mu(0)=z_\mu$ and
\begin{equation}
M=
\begin{pmatrix}
-1-i\epsilon&0&-\eta\gamma&\gamma\sqrt{1-\eta^2}\\
0&-1-i\epsilon&\gamma\sqrt{1-\eta^2}&\eta\gamma\\
-\eta\gamma&\gamma\sqrt{1-\eta^2}&-1-i\epsilon&0\\
\gamma\sqrt{1-\eta^2}&\eta\gamma&0&-1-i\epsilon
\end{pmatrix}\ ,
\label{21}
\end{equation}
while $\varphi(t)=\big(\langle z(t)|z(t)\rangle -\langle z|z\rangle\big)/2\eta$,
with $\langle z|z\rangle=\sum_{\mu=1}^4 z^*_\mu\, z_\mu$, the standard inner product.

Having found the proper time evolution for the fluctuation operators $a_i$, $b_i$, $i=1,2$, we want now
to study its physical properties, in particular, whether it is able to generate
quantum correlations among the two chains in the thermodynamical limit.

More specifically, we shall focus on two modes, $a_1$, that, involving only the fluctuation operators 
$\widetilde{X}_1$, $\widetilde{X}_2$, pertains to the first chain, 
and $b_1$, that instead involves $\widetilde{X}_5$, $\widetilde{X}_6$ and thus
belongs to the second chain ({\it cf.} the corresponding single site observables
in (\ref{15})). For generality, as initial state 
at time $t=\,0$ we shall take a squeezed thermal state, formed by the thermal state
$\widetilde{\omega}$ at inverse temperature $\beta$ introduced above, 
further squeezed with a common real parameter $r$ along 
the modes $a_1$ and $b_1$. In order to see whether at a later time $t$ these two modes
can get entangled by the action of the fluctuation dynamics in (\ref{20}),
we focus upon the properties of the reduced state obtained by tracing 
the full, four-modes state over the two modes $a_2$ and $b_2$.

Since the initial state is Gaussian and the dynamics Gaussian preserving,
all information concerning the time behaviour of the reduced, two-mode state is contained in its
$4\times 4$ covariant matrix $[\sigma]_{\mu\nu}=\frac{1}{2}\langle \{A_\mu,\, A_\nu^\dagger\}\rangle$,
where $A_\mu=(a_1,a_1^\dagger,b_1,b_1^\dagger)$ and $\langle\cdot\rangle$ means state average.
Applying the partial transposition criterion
to the case of Gaussian states \cite{Simon}, one can show that entanglement is present among the two
modes $a_1$ and $b_1$ if the smallest of the symplectic eigenvalues $\xi$ of the partially
transposed covariant matrix is negative.%
\footnote{Note that the reduced two-mode covariant matrix $\sigma$ is a principal minor
of the full covariant matrix $\Sigma$ involving all four modes $a_i$, $b_i$, $i=1,2$; therefore,
if the reduced state with covariance $\sigma$ is found entangled through 
the partial transposition criterion, {\it a fortiori} the same can be said for the
full four-mode state with covariance $\Sigma$.}
Actually, the logarithmic negativity, defined as \cite{Adesso}
\begin{equation}
E(t)={\rm max}\{0,\ -\log\xi\}\ ,
\label{22}
\end{equation}
gives a measure of the entanglement content of the state.

As clearly shown by the figures below, reporting the behavior in time of $E(t)$,
the dissipative, mesoscopic dynamics (\ref{20}) of the fluctuation algebra can indeed generate
quantum correlations starting from the chosen, completely separable initial state,
with a nonvanishing squeezing parameter $r$.
More specifically, the amount of the created entanglement increases 
as the dissipative parameter $\gamma$ gets larger ({\it cf.} Fig.1),
while it decreases and lasts for shorter times as the initial system 
temperature increases ({\it cf.} Fig.2). (In both figures, we have
taken the energy parameter $\epsilon$ to be one.)

In conclusion, we have shown that two independent, infinite spin-1/2 chains
can become entangled at the mesoscopic level through a purely dissipative mechanism,
via the action of a common heat bath.

\vskip 1cm

\begin{figure}[h]
\center{\includegraphics[width=0.63\textwidth]{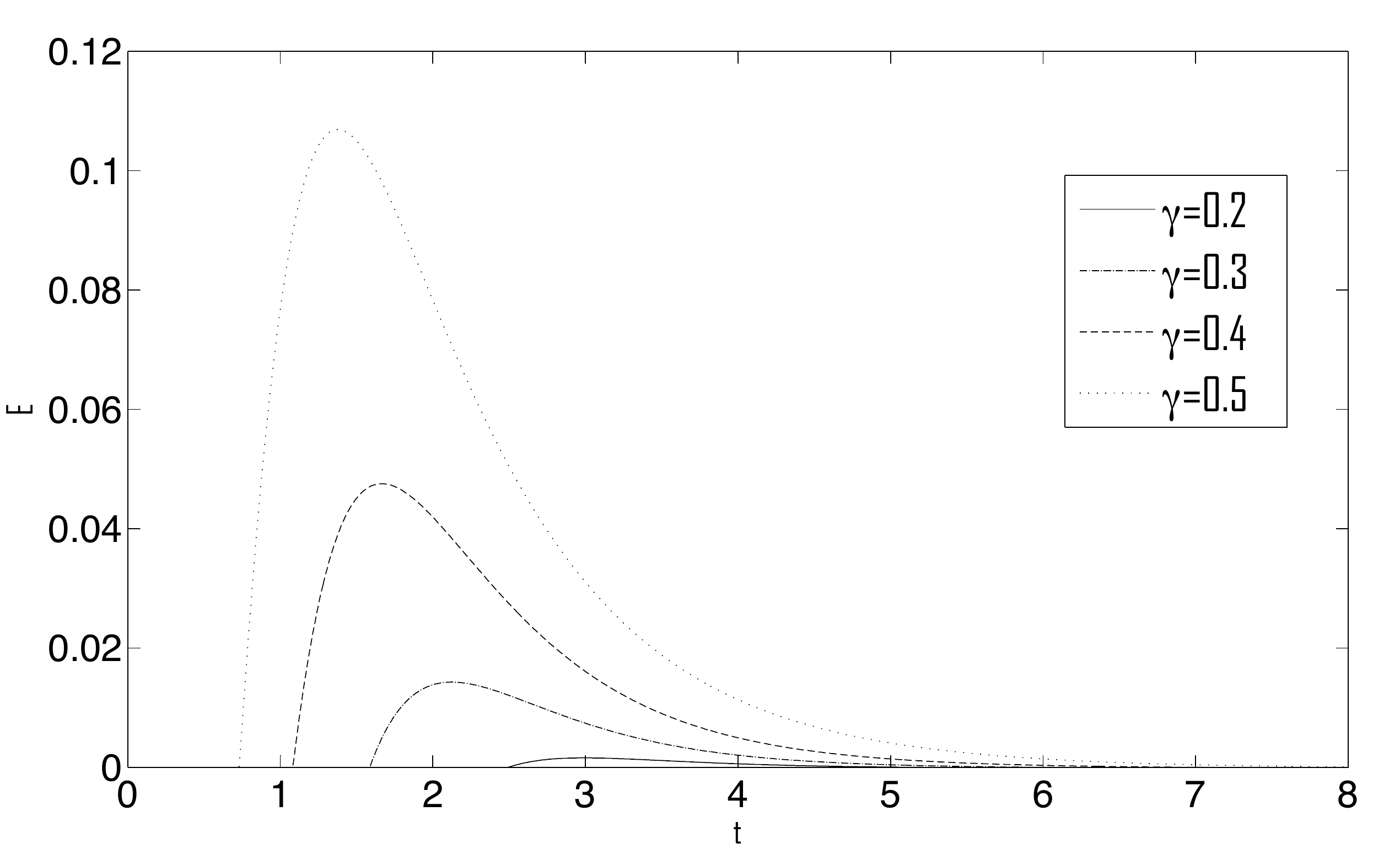}}
\caption{$E(t)$ for various values of the dissipative parameter $\gamma$,
with fixed temperature $T=0.1$ and squeezing $r=1$.}
\end{figure}

\vskip 2cm

\begin{figure}[h]
\center{\includegraphics[width=0.63\textwidth]{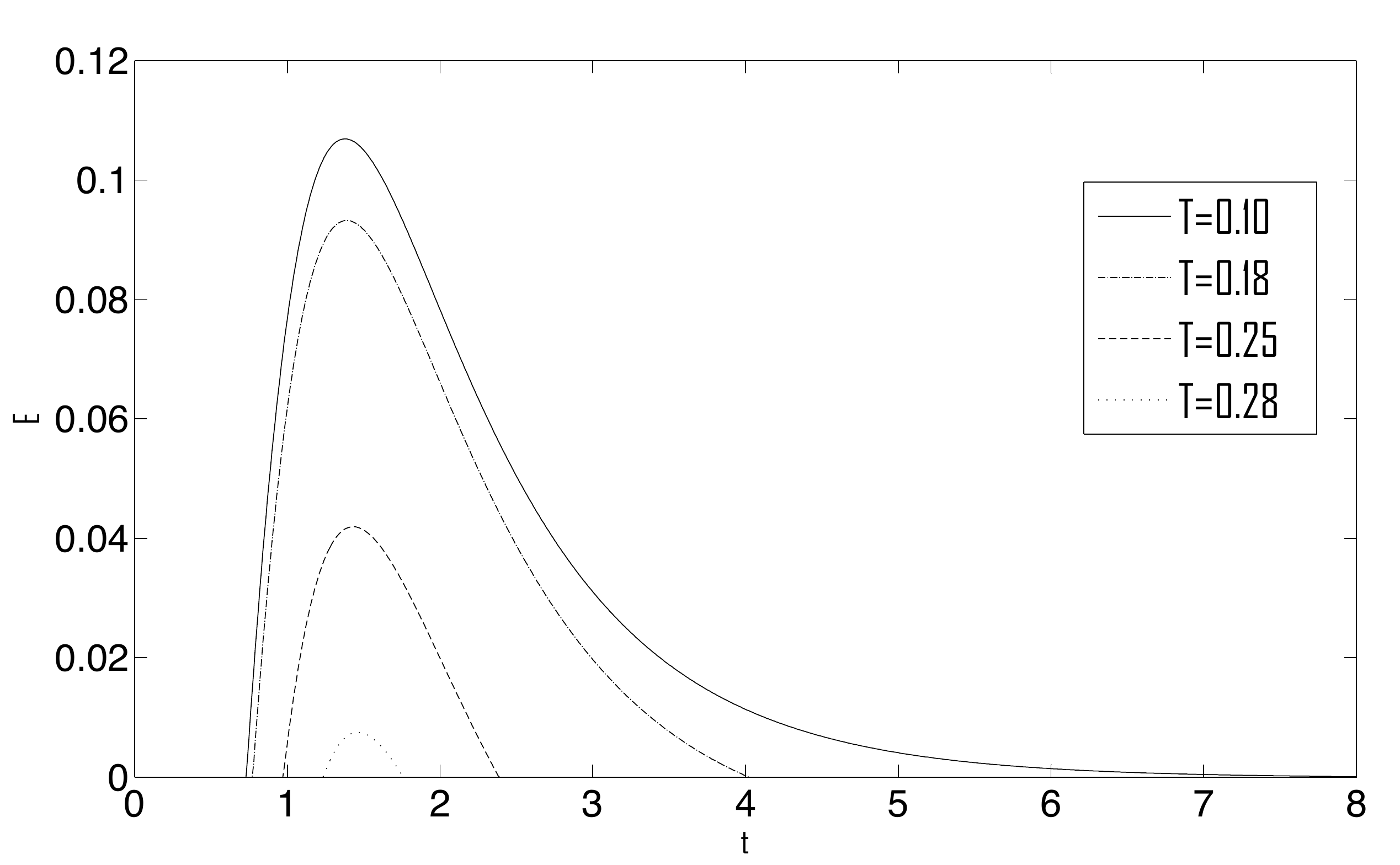}}
\caption{$E(t)$ for different values of the temperature,
with fixed dissipative, $\gamma=0.5$, and squeezing, $r=1$, parameters.}
\end{figure}

\vfill\eject

\end{document}